\def\aap{{A\&A}}
\def\aaps{{A\&AS}}

\def\apj{{ApJ}}
\def\apjs{{ApJS}}
\def\apjl{{ApJL}}
\def\mnras{{MNRAS}}

\def\FeIII{Fe\,{\sc iii}}
\def\FeIV{Fe\,{\sc iv}}
\def\Teff{$T_{\rm eff}$}
\def\logg{$\log g$}

\def\Rsun {$R_{\odot}$}
\def\Rsune {R_{\odot}}

\def\Rstar{$R_{\ast}$}

\def\Mdot{${\dot M}$}

\def\vinf {$v_{\rm \infty}$}
\def\vesc {$v_{\rm esc}$}

\def\Ha {H$_{\rm \alpha}$}

\def\kms {km\,s$^{-1}$}
\def\Mdu{$\cdot 10^{-6}\, {\rm M_{\odot}/yr}$}

\def\vrot{$v_{\rm rot}$}
\def\vsini{$v\,\sin i$}

\def\beq{\begin{equation}}
\def\eeq{\end{equation}}
\def\beqa{\begin{eqnarray}}
\def\eeqa{\end{eqnarray}}

\def \Lsune {L_{\odot}}

\def \Rstare {R_\star}

\def \Teffe {T_{\rm eff}}

\def \vinfe {v_\infty}
\def \vesce {v_{\rm esc}}
\def \Mdote {\dot M}

\documentclass{iaus}
\usepackage{graphicx, natbib}

\title[OB-stars as extreme condition test beds] 
{OB-stars as extreme condition test beds}

\author[J. Puls, J.O. Sundqvist, \& J.G. Rivero Gonz\'alez]   
{Joachim Puls$^1$, Jon O. Sundqvist$^1$,
 \and Jorge G. Rivero Gonz\'alez$^1$}

\affiliation{$^1$ Universit\"atssternwarte der 
Ludwig-Maximilians-Universit\"at M\"unchen \\ Scheinerstr. 1, 
D-81679 M\"unchen, Germany \\ email: {\tt uh101aw@usm.uni-muenchen.de} (J.P.)} 

\pubyear{2010}
\volume{272}  
\pagerange{119--126}
\jname{Active OB stars: structure, evolution, mass loss and critical limits}
\editors{C. Neiner, G. Wade, G. Meynet \& G. Peters, eds.}
\begin{document}

\maketitle

\begin{abstract} 
Massive stars are inherently extreme objects, in terms of radiation, mass
loss, rotation, and sometimes also magnetic fields. Concentrating on a
(personally biased) subset of processes related to pulsations, rapid
rotation and its interplay with mass-loss, and the bi-stability mechanism, we
will discuss how active (and normal) OB stars can serve as appropriate
laboratories to provide further clues. 
\end{abstract}

\firstsection 
\section{Introduction} Massive stars are inherently extreme objects, in
terms of radiation, mass loss, rotation and sometimes also magnetic fields.
Thus, they can serve as test beds for extreme conditions and corresponding
theoretical predictions.  Such tests are, e.g., particularly important for
our understanding of the (very massive) First Stars and for the physics of
fast rotation in massive stars, which is a key ingredient in the collapsar
model of long Gamma Ray Bursts. In this review we discuss how a variety of
physical processes present in massive stars can affect both their stellar
photospheres and/or winds, and how active (and normal) OB stars can be, and
are, used as appropriate laboratories to provide further clues. In the
following, we concentrate on a (personally biased) subset of processes
related to pulsations (Sect. 2), rapid rotation and its interplay with
mass-loss (Sect. 3), and mass-loss itself, particularly on the bi-stability
mechanism (Sect. 4).

\section{Pulsations} 

\begin{figure}[t]
\begin{center}
 \includegraphics[width=10cm]{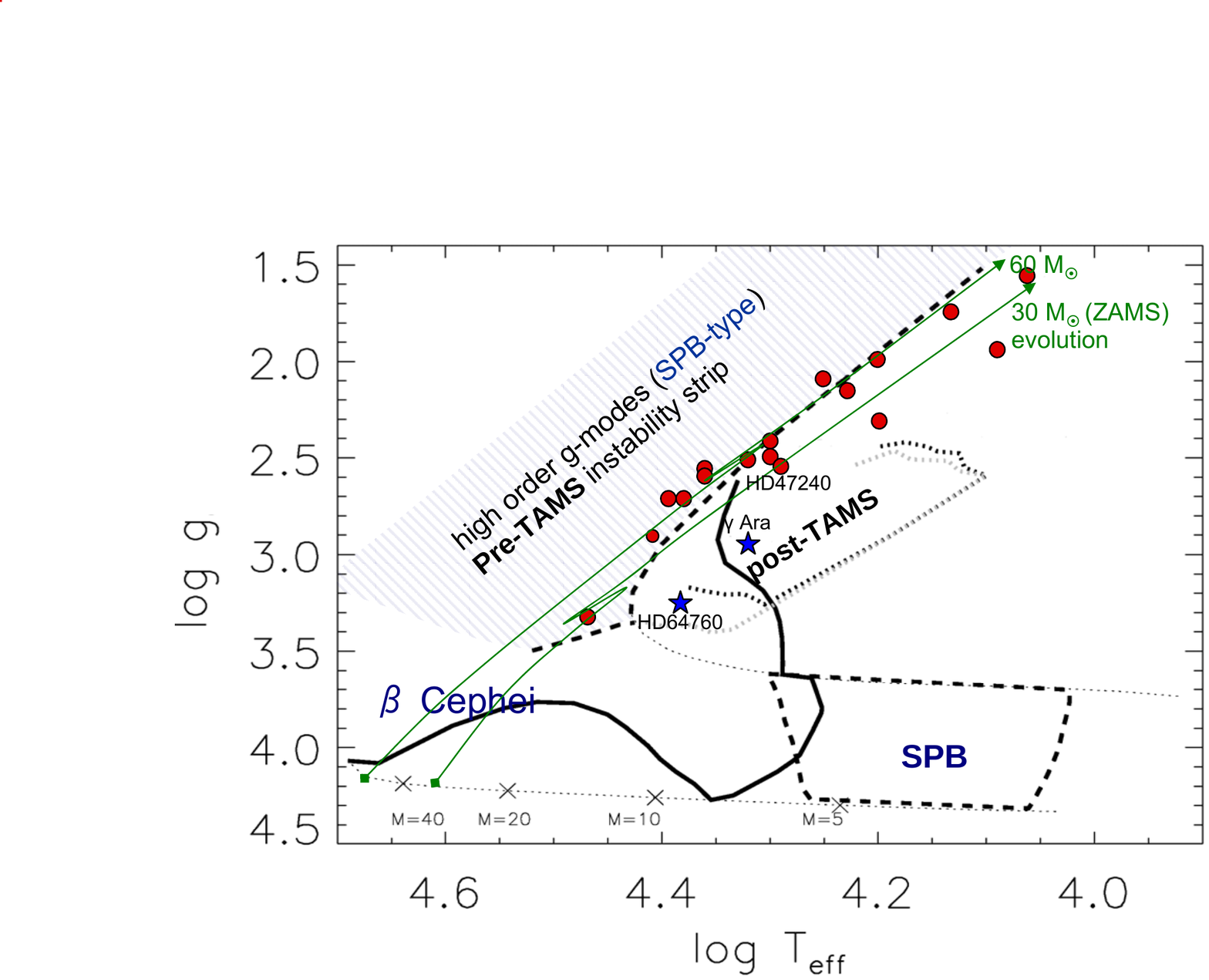} 
\vspace*{-0.2 cm}
\caption{$\log$ \Teff-\logg\ diagram for hot massive stars. Indicated are the
instability regions for $\beta$ Cep stars (solid bold), SPBs close to the
ZAMS (dashed bold) and SPBs of evolved type as predicted by 
\citet{Pamyatnykh99} (pre-TAMS, blue hatched) and \citet{Saio06} (post-TAMS, 
dotted). 
Red dots correspond to the positions of
slowly pulsating B-supergiants from the sample by \citet{Waelkens98} as
derived by \citet{Lefever07}, and blue asterisks are two newly suggested
g-mode pulsators from the same work. See text for further details. Adapted
from \citet{Lefever07}.}
\label{fig1}
\end{center}
\end{figure}

\noindent
{\bf Pulsating B-supergiants.}
Well outside the instability strips of $\beta$ Cep and slowly pulsating 
B-stars (SPB),
\citet{Waelkens98} via {\sc hipparcos} detected 29 periodically variable
B-{\it supergiants}. A corresponding instability region had not been
predicted at that time. Meanwhile, however, \citet{Pamyatnykh99} and \citet[
see also this volume]{Saio06} identified such regions for pre-TAMS and
post-TAMs objects, respectively, with SPB-type of oscillations (high order
g-modes). These regions are indicated in Fig.~\ref{fig1}, together with
results from quantitative spectroscopy by \citet{Lefever07}, for those of
the above 29 supergiants with sufficient spectral information. Obviously,
most of these objects are located very close to the high gravity limit of
the predicted pre-TAMS or within the predicted post-TAMS instability strips
for evolved stars. Together with their multi-periodic behaviour, this
strongly suggests that these objects are opacity-driven non-radial pulsators
(NRPs), and thus are ideal {\bf test beds} for asteroseismologic studies of
evolved massive stars. Note that \citet{Lefever07} found additional
periodically variable objects not known to be pulsators so far, and
suggested, from their pulsational behaviour and their positions, that these
objects are g-mode pulsators as well. Two of them, HD\,64760 (B0.5\,Ib) and
$\gamma$~Ara (= HD\,157246, B1\,Ib), are explicitly indicated in
Fig.~\ref{fig1}, together with HD\,47240 (B1\,Ib) from the original sample
by Waelkens et al., and will be referred to later on. 

\smallskip \noindent {\bf Macroturbulence.} A present key problem in
atmospheric diagnostics by high resolution spectroscopy is the finding that
the line-profiles from (at least) late O- and B-supergiants display
substantial extra-broadening (in addition to the well-known effects from
rotation etc.), which has been termed `macro-turbulence' (for details and
references, see \citealt{SimonDiaz10}). This extra-broadening can be
simulated by allowing for a {\it supersonic} Gaussian or quasi-Gaussian
velocity distribution in photospheric regions, which is difficult to justify
physically. Recently, however, \citet{Aerts09} showed that such
extra-broadening can be reproduced from the {\it collective} effect of
low-amplitude g-mode oscillations. First hints that this scenario might be
realistic have been found by \citet{SimonDiaz10}, from a tight observed
correlation between the peak-to-peak amplitudes of velocity moments measured
from (variable) photospheric profiles of B-supergiants and the derived
macro-turbulent broadening. Given the ubiquity of macro-turbulence in hot
massive stars, this tentatively suggests that a large fraction of 
OB-stars are non-radial pulsators (see also
Fullerton, this volume).

\smallskip \noindent {\bf Triggering of structure/clump formation.} With
respect to their stellar winds, pulsations in massive stars might be
responsible for inducing large-scale structures, such as co-rotating
interaction regions (CIRs, see Fullerton, this volume), and, particularly,
for triggering the formation of clumps: To reproduce the observed X-ray
emission from hot stellar winds, $L_{\rm x} \approx 10^{-7} L_{\rm bol}$,
the line-driven (or deshadowing) instability related to radiative line
driving needs to be excited by deep-seated photospheric disturbances of a
multitude of frequencies (NRPs?), giving rise to strong clump-clump
collisions and consequently strong shocks \citep{Feld97a, Feld97b}. For
models with self-excited instability alone, the predicted X-ray emission is
much too weak. Moreover, such perturbations might be responsible for
triggering the on-set of {\it deep-seated} wind-clumping (Sect.~4), as
implied from various diagnostics (e.g., \citealt{Bouret05, Puls06,
Sundqvist11}).

\smallskip \noindent {\bf Strange mode oscillations.} In addition to
`conventional' pulsations, another class of quasi-periodic, {\it dynamical}
instabilities are predicted to occur in the envelopes of luminous stars with
large $L/M > 10^3$. These are the so-called strange-mode oscillations (for
details and references, see Saio and Chen\'e, this volume), which should be
particularly strong in WR-stars and might even help to initiate their winds
(e.g., \citealt{Wende08}). So far, there is no direct evidence of these
predictions, though the strongest amplitudes of optical lpv in O-stars are
located within the region of predicted strange mode oscillations
\citep{Fullerton96}, and at least for one WR star such oscillations might
actually have been observed (see Chen\'e, this volume).  Alternative {\bf
test beds} to check the reality of strange mode oscillations might be late
B-/early A-supergiants (as suggested by Puls, Glatzel, \& Aerts as targets
for the micro-satellite {\sc brite}), since, in comparison to WRs, these
objects have less dense winds and `convenient' frequencies (on the order of
a few to tens of days), with predicted amplitudes of 0.1 mag (W.~Glatzel,
priv. comm.). Indeed, the {\sc corot} observations of the late B-supergiant
HD\,50064 \citep{Aerts10} showed a period of 37 days, with a sudden
amplitude change by a factor of 1.6. Together with other evidence (variable
\Mdot\ etc.), Aerts et al. tentatively interpreted this finding as the
result of a strange mode oscillation.

\section{Rapid rotation}

\noindent
{\bf Photospheric deformation and gravity darkening.} Rapid rotation affects
the stellar photosphere in (at least) two ways. First, it becomes deformed,
with $R_{\rm eq}/R_{\rm pole}$ = 1.5 at critical rotation (using a Roche
model with point mass distribution, see Zhao, this volume, and
\citealt{CO95} for details and references). The first observational {\bf
test bed} which confirmed the basic effect was the brightest Be star
known, Achernar = $\alpha$ Eri (VLTI observations by \citealt{deSouza03}).

\begin{figure}[t]
\begin{center}
\begin{minipage}{6.5cm}
\includegraphics[angle=90,width=6.5cm]{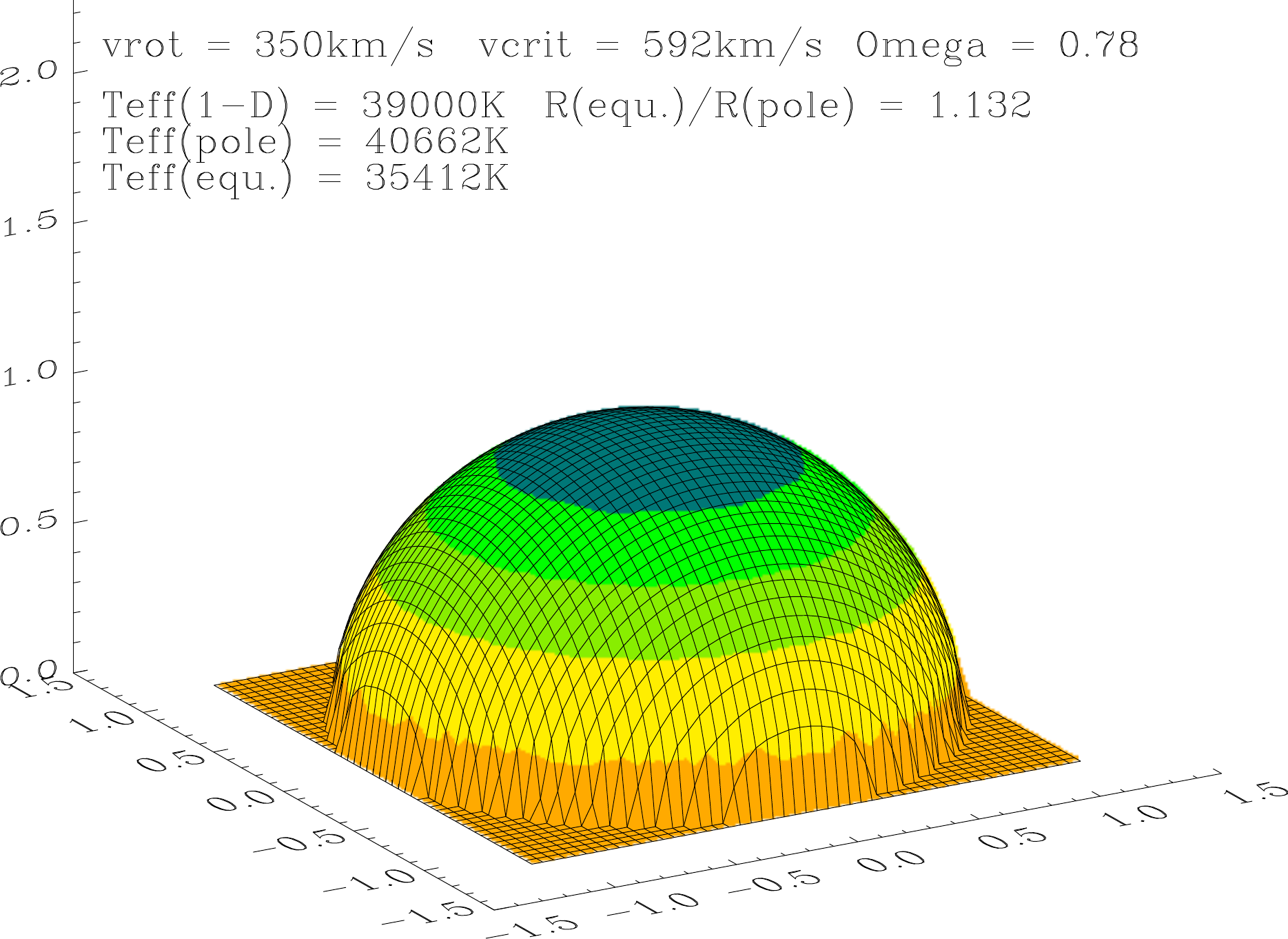} 
\end{minipage}
\hfill
\begin{minipage}{6.5cm}
\includegraphics[angle=90,width=6.5cm]{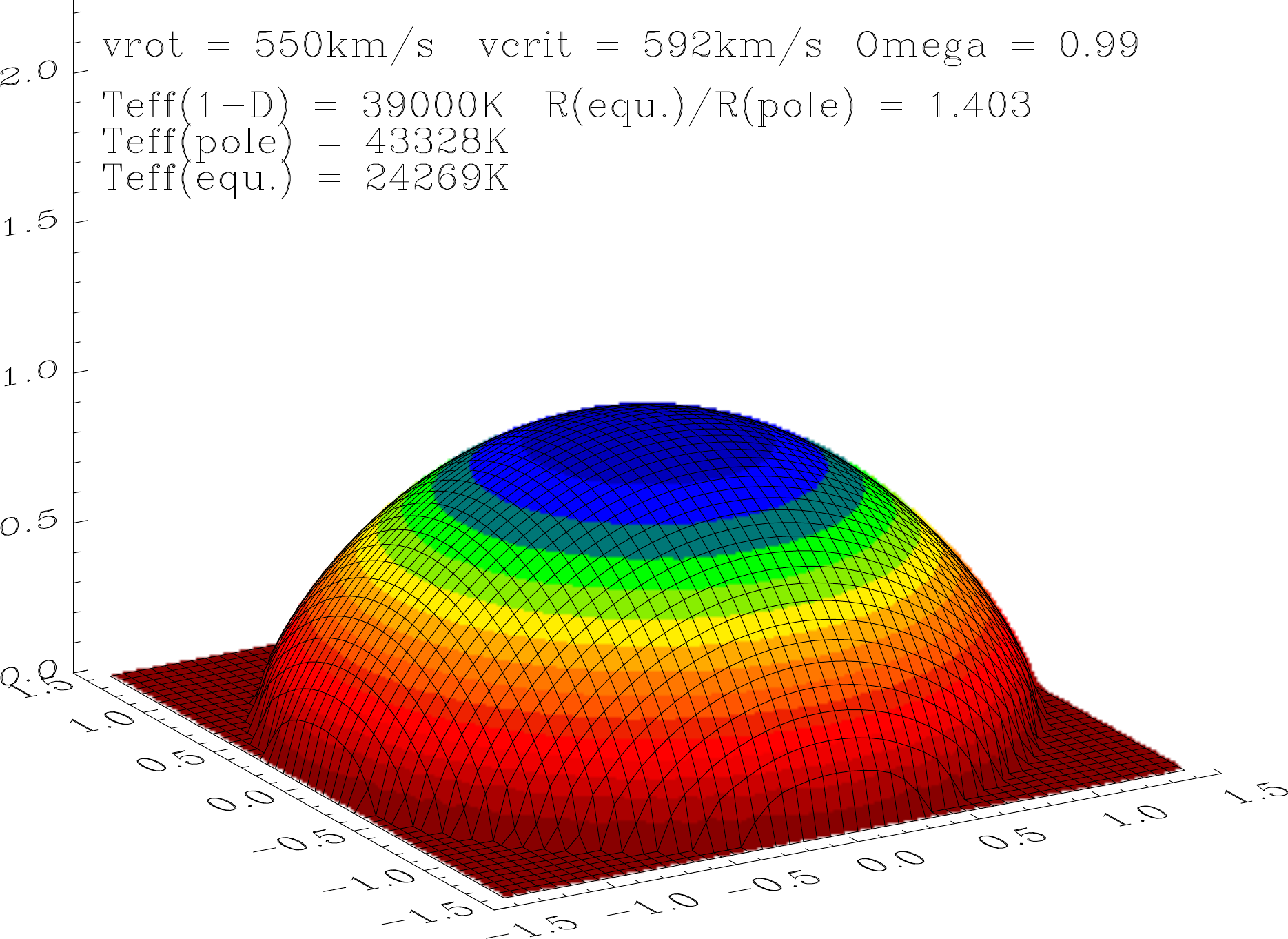} 
\end{minipage}
%
\caption{Predicted photospheric deformation and gravity darkening for a star
similar to $\zeta$~Pup (1-D values: \Teff=39~kK, \Rstar=19\Rsun, \logg=3.6)
but rotating with 78\% (left) and 99\% (right) of its critical angular
velocity. Both figures are on the same scale, with identical color coding
for \Teff($\vartheta$). The ratios $R_{\rm eq}/R_{\rm pole}$ and the
effective temperatures at the hotter pole and cooler equator are indicated
within the figures.}
\label{fig2}
\end{center}
\end{figure}

The second effect is gravity darkening, first suggested by
\citet{vonZeipel24}, who assumed rotational laws that can be derived from a
potential, e.g., uniform or cylindrical. An important extension was provided
by \citet{Maeder99}, who considered the more realistic case of shellular
rotation in radiative envelopes, where the angular velocity is assumed to be
constant on horizontal surfaces \citep{Zahn92}. In result, the photospheric
flux is proportional to the {\it effective} gravity, $\vec{F} \propto
\vec{g}_{\rm eff} (1+\zeta(\vartheta))$, with $|\zeta(\vartheta)| < 0.1$ in
most cases and $\zeta$ = 0 in the original von Zeipel case. The effective
gravity is the vector sum of gravitational and centrifugal acceleration,
$\vec{g}_{\rm eff} = \vec{g}_{\rm grav} + \vec{g}_{\rm cent}$, i.e., lower
at the equator than at the pole, with $\vec{g}_{\rm eff}$(pole) =
$\vec{g}_{\rm grav}$. Note that here $\vec{g}_{\rm eff}$ is {\it independent of
the radiative acceleration}! Neglecting $\zeta(\vartheta)$, for radiative
envelopes we obtain $\Teffe(\vartheta) \propto g_{{\rm eff},\perp}^{1/4}$,
i.e., \Teff\ decreases towards the equator, in dependence of the normal
component of $\vec{g}_{\rm eff}$.

Both effects are demonstrated in Fig.~\ref{fig2}, for a typical O-supergiant
rotating close and very close to critical rotation. Deformation and gravity 
darkening become significant only for rotational speeds higher than roughly
70\% of the critical one! {\bf Test beds} to check both effects are
discussed by Zhao (this volume; see also the summary provided by
\citealt{vanBelle10}). 

\smallskip \noindent {\bf Rapid rotation and winds.} Standard line-driven
wind theory (for a recent review and references, see \citealt{Puls08})
predicts that the mass-loss rate of a non (slowly) rotating star scales as
%
\[
\Mdote \propto (N_{\rm eff} L)^{1/\alpha} \bigl(g_{\rm grav}
\Rstare^2(1-\Gamma)\bigr)^{1-1/\alpha},
\]
%
with $N_{\rm eff}$ the effective number of driving lines (proportional to
the force-multiplier parameter $k$, \citealt{CAK75}, CAK), CAK parameter
$\alpha$ (corresponding to the steepness of the line-strength distribution
function), and Eddington factor $\Gamma$. Accounting for rotation (and
$\Gamma$ not too large), we find that the mass-loss rate depends on
co-latitude~$\theta$,
\beqa
\Mdote(\theta) & \propto & 
\bigl(N_{\rm eff}(\theta) F(\theta)\Rstare^2(\theta)\bigr)^{1/\alpha(\theta)} 
\bigl(g_{\rm eff}(\theta)
\Rstare^2(\theta)(1-\Gamma)\bigr)^{1-1/\alpha(\theta)} \nonumber \\
& \stackrel{\rm von\, Zeipel}{\propto} & 
\bigl(N_{\rm eff}(\theta)\bigr)^{1/\alpha(\theta)} 
g_{\rm eff}(\theta)\Rstare^2(\theta)
\label{mdot_rot}
\eeqa
(cf. \citealt{Owo98}). This expression renders two possibilities. i) If the
ionization equilibrium is rather constant w.r.t. $\theta$ (as is the case
for O-stars), we obtain 
a {\it prolate} wind structure, since $g_{\rm eff}(\theta)$ is largest at the
pole. This is the $g_{\rm eff}$-effect, see \citet{Owo98, Maeder99, MM00}.
ii) {\it If}, on the other hand, the ionization equilibrium were strongly
dependent on $\theta$, this would
imply 
an {\it oblate} wind structure if the increase of $N_{\rm eff}$ and the
decrease of $\alpha$ towards the equator (as a consequence of decreasing
ionization) could overcompensate the decrease of
$g_{\rm eff}$. Such a situation (the $\kappa$-effect, see \citealt{Maeder99,
MM00}) {\it might} occur in B-supergiants (but see Sect.~4). Note, however,
that {\it no} thin disk can be formed by this process alone. Note also that
self-consistent 2-D hydro/NLTE calculations (though somewhat simplified) for
rapidly rotating B-stars around \Teff = 20~kK (i.e, just in the region where
the $\kappa$ effect might be expected) by \citet{PP00} still resulted in a
prolate wind structure, since the ionization effects turned out to be
only moderate.

Of course, these predictions need to be checked observationally,
particularly when considering their importance regarding stellar evolution
(e.g., a pronounced polar mass loss would lead to less loss of angular
momentum), and with respect to mass-loss diagnostics (when do we need 2-D
models?). To our knowledge, clear observational evidence for aspherical
winds is still missing.\footnote{The polar wind structures claimed for the
Be-stars Achernar \citep{Kervella06} and $\alpha$ Ara \citep{Meilland07}
from NIR interferometry still need to be confirmed, given that -
as discussed during this conference - for such low mass-loss rates the
IR-photosphere is very close to the optical one (in other words, the
IR-excess from the wind is very low).} So, what are potential {\bf test
beds}?

\begin{figure}[t]
\begin{center}
 \includegraphics[width=10cm]{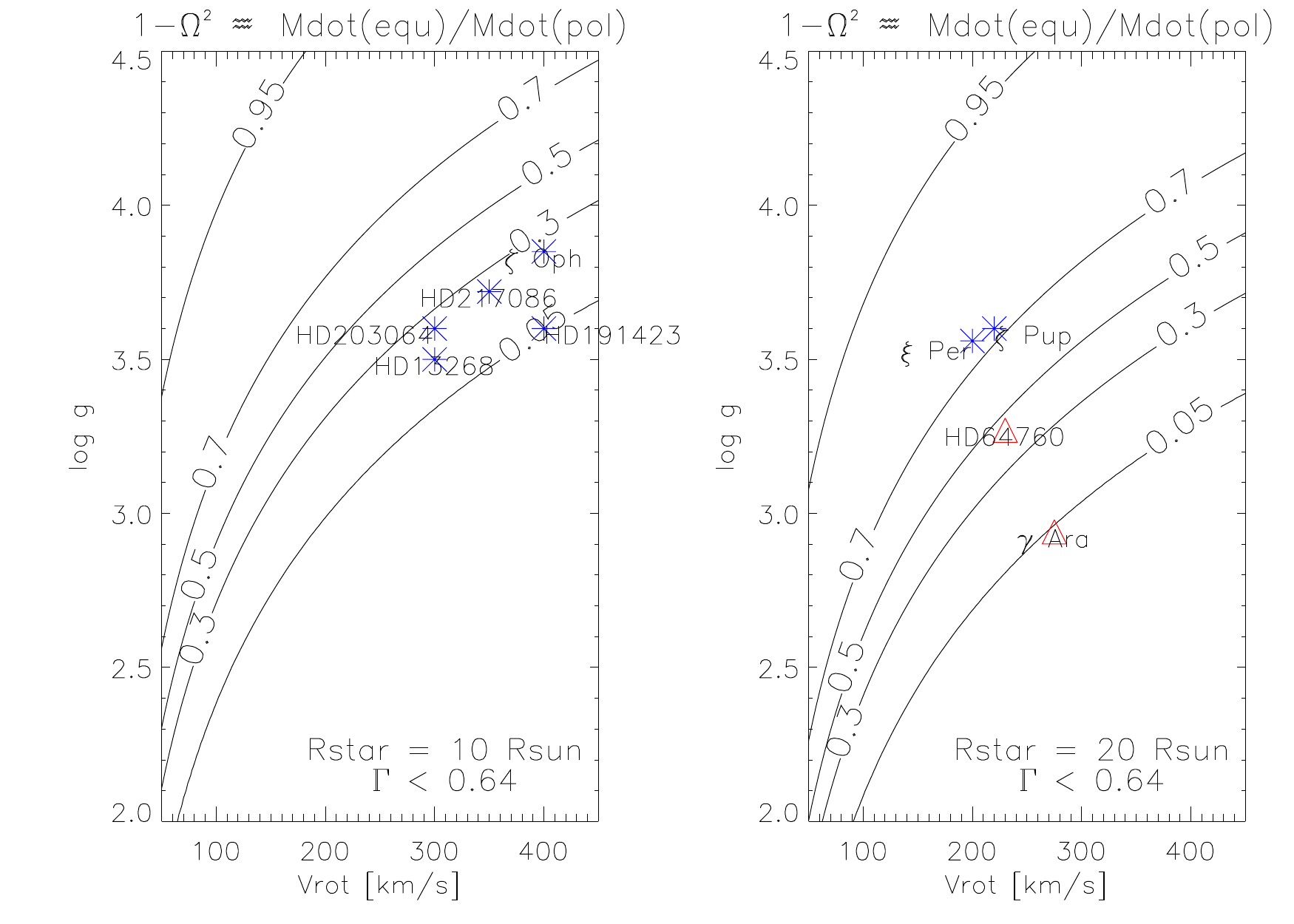} 
\vspace*{-0.3 cm}
\caption{Iso-contours of $1-\Omega^2$ $\approx$ \Mdot(eq)/\Mdot(pole) as a
function of \vrot\ and \logg, for typical dwarfs (left) and supergiants
(right). Overplotted are the positions of some rapidly rotating Galactic
O-stars (asterisks) and B-supergiants (triangles), assuming a minimum \vrot
= \vsini.} \label{fig3}
\end{center}
\end{figure}
\begin{figure}[b]
\vspace*{-0.5 cm}
\begin{center}
\begin{minipage}{6.5cm}
 \includegraphics[angle=180,width=6.5cm]{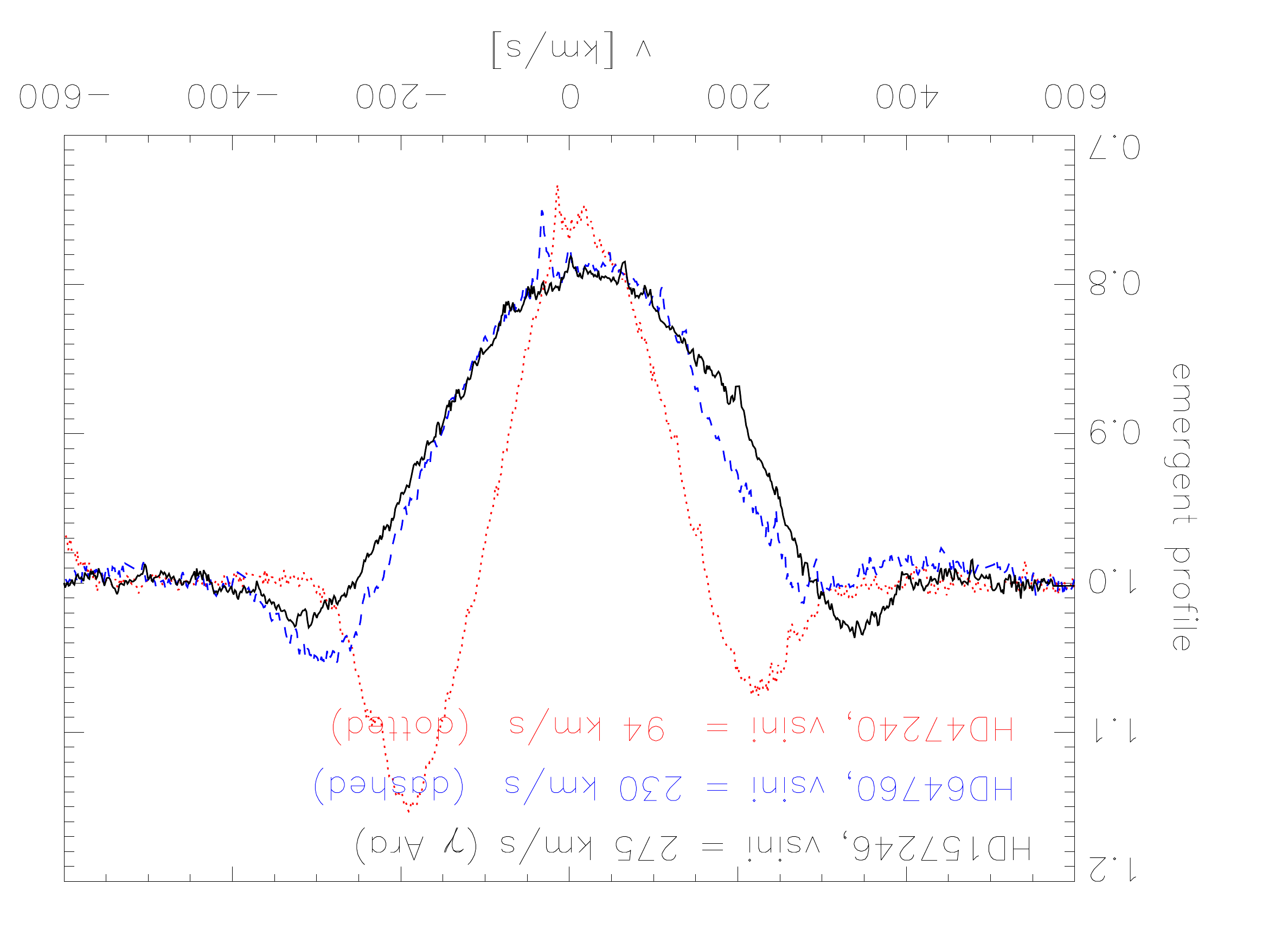} 
\end{minipage}
\begin{minipage}{5cm}
 \includegraphics[width=5cm]{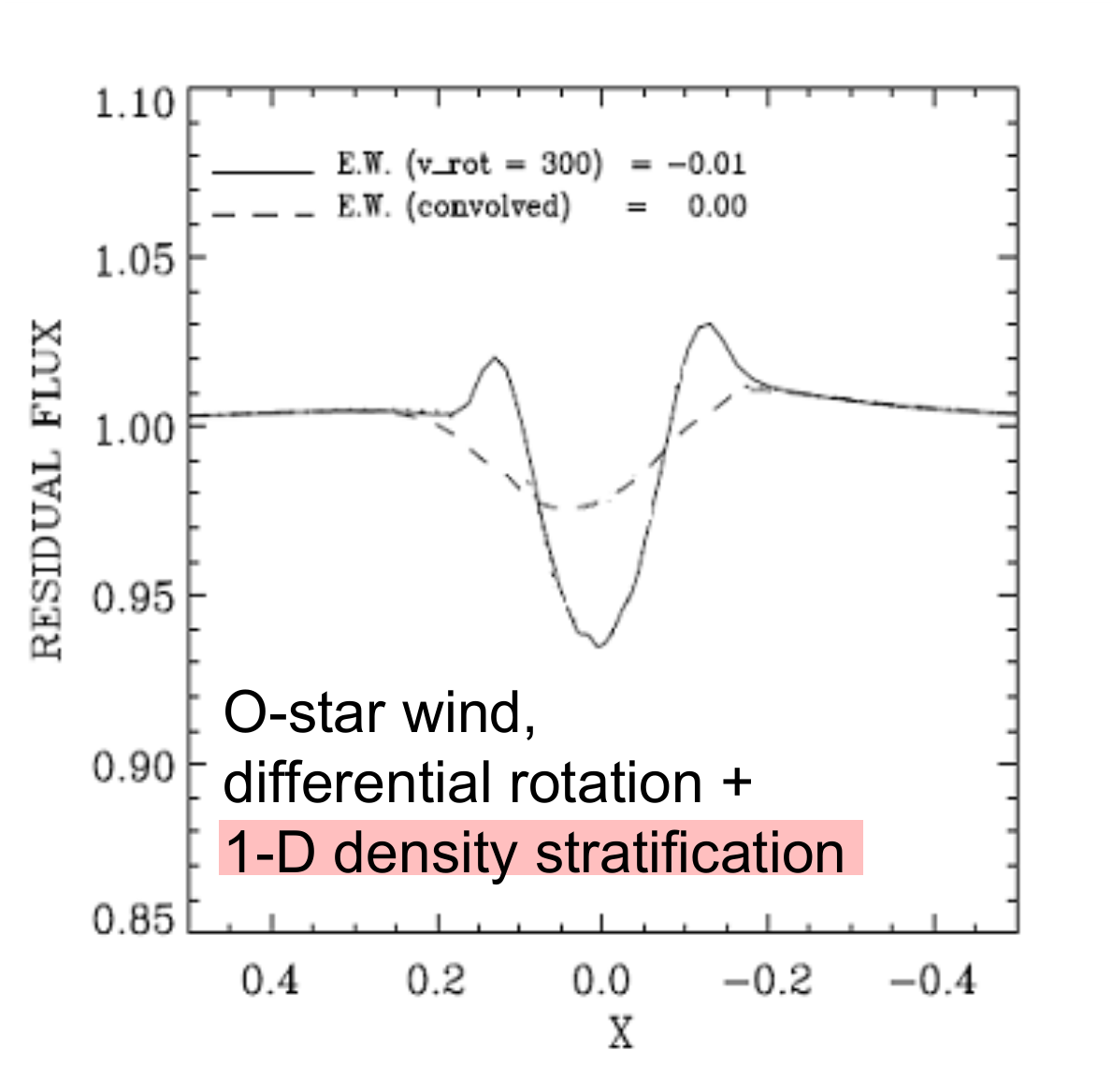} 
\end{minipage}
\vspace*{-0.3 cm}
\caption
{Left: \Ha\ line profiles for three rapidly rotating B-supergiants (spectra
and data from \citealt{Lefever07}). Right: Theoretical \Ha\ line profiles,
for an O-star wind with {\it spherical} density stratification. Dashed:
Profile convolved with a rotation profile of width 300~\kms. Solid: 2-D line
transfer allowing for differential rotation, \vrot $\propto 1/r$. Adapted
from \citet{PP96}.}
\label{fig4}
\end{center}
\end{figure}

In Fig.~\ref{fig3} we have plotted theoretical iso-contours of $1-\Omega^2$
$\approx$ \Mdot(eq)/\Mdot(pole) (from Eq.~\ref{mdot_rot} and assuming 
$N_{\rm eff}$ and $\alpha$ to be constant; $g_{\rm eff} \approx$ $g_{\rm
grav}(1-\Omega^2)$ and $\Omega = \omega/\omega_{\rm crit}$) as a function of
\vrot\ and \logg, for typical dwarfs (left) and supergiants (right).
Overplotted are the locations of well-known Galactic rapid rotators,
for a minimum value of \vrot = \vsini\ (data from \citealt{Repo04} and
\citealt{Lefever07}). Asterisks denote O-stars, triangles
B-supergiants. All O-dwarfs/giants (left) are predicted to have a 
significant mass-loss contrast, below 0.3.  Unfortunately, their (average)
mass-loss rate is too low to lead to substantial effects in the optical
wind-lines and the IR continuum, though UV-spectra should be
affected by deviations from spherical symmetry. For the fast rotating
O-supergiants, on the other hand, the predicted effect is rather small, so
nothing might be visible. \citet{Vink09}, using linear \Ha\
spectro-polarimetry, conclude that most winds from rapidly rotating O-stars
are spherically symmetric. For the two rapidly rotating B-supergiants,
HD\,64770 and particularly $\gamma$~Ara, the situation is more promising,
and they might be used as {\bf test beds} to check the impact of rotation on the
global wind topology. Remember that HD\,64760 (see also Sect.~2) is one of
the best studied objects in the UV (thanks to the IUE mega-campaign,
\citealt{Massa95}) - with the detection of CIRs and `PAMS' (see Fullerton,
this volume), both presumably related to its non-radial pulsations --, and has
also been studied in the optical to clarify the interaction between NRPs and
CIRs \citep{Kaufer06}.

Fig.~\ref{fig4} (left) displays the corresponding \Ha\ line profiles, for
the above two rapidly rotating B-supergiants and for HD\,47240 (see also
Sect.~2), with a somewhat lower \vsini. At first
glance, these profiles might indicate the presence of a disk or an oblate
wind (e.g., the $\kappa$-effect from above), but this is not necessarily the
case. As shown in Fig.~\ref{fig4} (right), even a {\it spherical} wind can
give rise to double-peaked profiles, when accounting for the wind's
differential rotation (due to the so-called resonance-zone effect, 
\citealt{PP96}). Note that in this case the profile only depends on the product
\vsini\ and not on the individual factors. For a 2-D density stratification,
however, the profiles will look different for a prolate or oblate topology,
and will depend on the individual values of \vrot\ and $\sin i$ as well,
which might induce a certain dichotomy. Interestingly, UV spectroscopy (via
IUE) of $\gamma$ Ara by \citet{Prinja97} gave indications for a {\it
prolate} geometry, mainly because of missing or weak emission peaks in the P
Cygni profiles.

\begin{figure}[b]
\vspace*{-0.3 cm}
\begin{center}
 \includegraphics[angle=180,width=9.0cm]{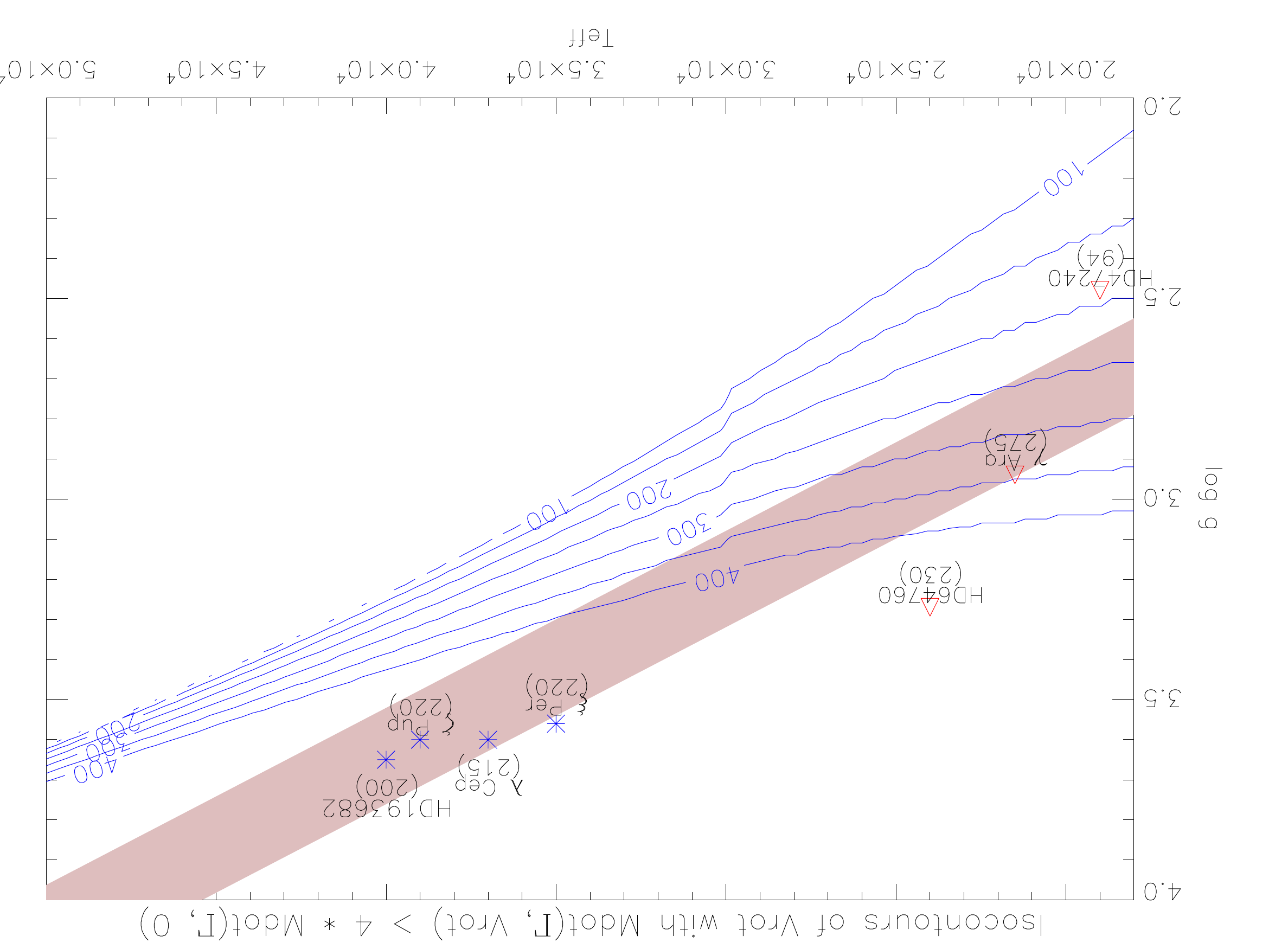} 
\vspace*{-0.3 cm}
\caption
{Iso-contours of \vrot\ for which the total mass-loss rate is
predicted to become increased by a factor of four, compared to the
non-rotating case, as a function of \Teff\ and \logg, together with the
positions and \vsini-values of some rapidly rotating O- and
B-supergiants (asterisks and triangles, respectively). See text.}
\label{fig5}
\end{center}
\end{figure}

\smallskip \noindent {\bf The $\Omega\Gamma$ limit.} An interesting question
is what happens if a star is rapidly rotating {\it and} close to the
Eddington limit. After a controversial discussion \citep{Langer97,
Glatzel98}, \citet{MM00} were able to solve this problem in an elegant way.
For the following discussion, it is only important to note that the {\it
total} acceleration due to gravity, centrifugal forces, and radiation
pressure gradients can be expressed as $\vec{g_{\rm tot}} = \vec{g_{\rm
eff}}(1 - \Gamma_\Omega)$, where the effective gravity remains defined as
previously, and $\Gamma_\Omega/\Gamma > 1$ is a function of \vrot/$v_{\rm
crit}$. Consequently, the total acceleration can become zero before the
nominal Eddington limit is reached, and this new limit is called the
$\Omega\Gamma$ limit. As shown by \citet{MM00}, the combination of rapid
rotation and large $\Gamma$ can affect the total (polar-angle integrated)
mass-loss rate from a radiation driven wind considerably,  
%
\[
\frac{\Mdote(\mbox{rotating})}{\Mdote(\mbox{non-rotating})}
\approx \Bigl(\frac{1-\Gamma}{\Gamma/\Gamma_\Omega - 
\Gamma}\Bigr)^{\frac{1}{\alpha} -1} 
\left\{ \begin{array}{ll} = O(1) & \mbox{for not too fast rotation and low
$\Gamma$} 
\\ \gg 1 & \mbox{for fast rotation and considerable $\Gamma$} \\
\mbox{but:} & \mbox{max. \Mdot\ limited because $L$ limited}
\end{array} \right.
\]
since $\alpha$ is on the order of 0.4 {\ldots} 0.6. To identify potential
{\bf test beds} to check this important prediction, in Fig.~\ref{fig5} we
have plotted the iso-contours of the \vrot\ required for a significantly
increased mass-loss rate, as a function of \Teff\ and \logg. (A factor of
four compared to the non-rotating case was chosen to allow for an easy
observational check.) The red shaded region comprises the approximate 
locations of Galactic OB-supergiants. Overplotted are the positions of some
rapidly rotating supergiants, O-types (asterisks) and B-types (triangles). 
The numbers in brackets are the observed \vsini. Again,
O-supergiants are not suited as test beds, since they would need to rotate
much faster than 400~\kms\ to show the required increase in \Mdot.
Interestingly, however, at least $\gamma$ Ara (and maybe HD\,47240 - if its
$\sin i$ were 0.5) are located at the `right' position and might be worth
being investigated in detail, e.g., by means of interferometry (see
Chesneau, this volume) in combination with 2-D NLTE modeling
\citep{Georgiev06}. The outcome of such investigations will be of 
particular relevance for stellar evolution with rapid rotation, especially
in the early Universe (see Ekstr\"om, this volume).

\section{Mass loss}

\noindent 
As we have seen, rapidly rotating B-supergiants are ideal {\bf test beds} to
check a number of theoretical predictions. Unfortunately, however, there
exists only few such objects, since there is a rapid drop of rotation
below \Teff $\approx$ 20~kK, as is obvious from the distribution of \vsini\
(e.g., \citealt{Howarth97}). In a recent letter,
\citet{Vink10} tried to explain this finding based on two alternative
scenarios (see also Langer, this volume). In {\it scenario I}, the low rotation
rates of B-supergiants are suggested to be caused by braking due to an
increased mass loss for \Teff $<$ 25~kK, where this increased mass loss
should be due to the so-called bi-stability jump. Vink et al. termed this
process `bi-stability braking'.  

\begin{figure}[t]
\begin{center}
 \includegraphics[width=7.7cm]{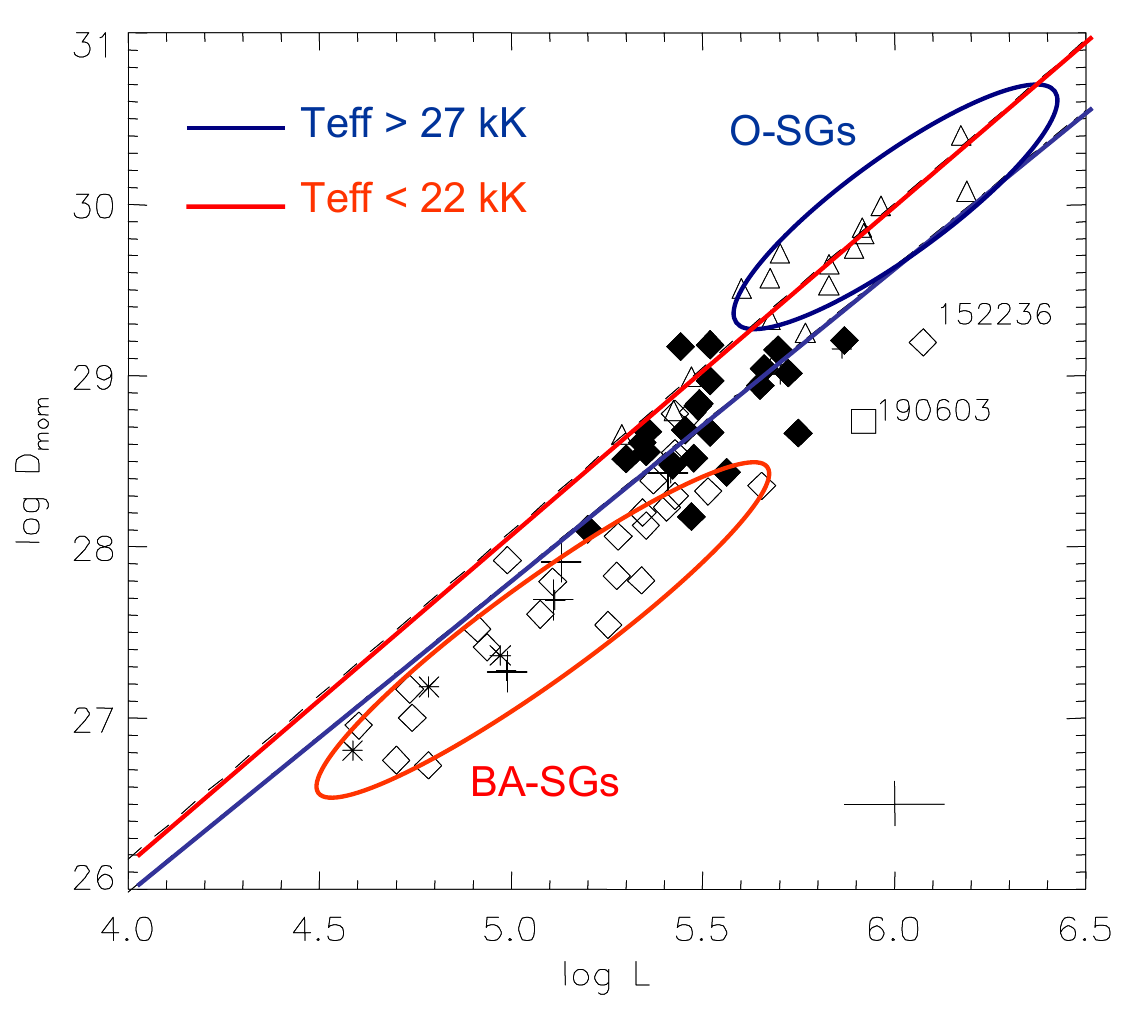} 
\vspace*{-0.3 cm}
\caption{Modified wind-momenta for Galactic O- and B-supergiants. Dashed
(red) and solid (blue) lines represent the predictions by \citet{Vink00} for
objects below and above 23~kK, respectively. Triangles are O-supergiants
(\Teff\ $>$ 27~kK), filled diamonds early B-sgs with \Teff\ $>$ 22~kK and
open triangles B-supergiants below this value. The cross in the lower right
displays the typical error bars. Adapted from \citet{Markova08}.} 
\label{fig6}
\end{center}
\end{figure}

\smallskip \noindent {\bf The bi-stability jump} itself has often been 
discussed and referred to in the literature, and goes back to findings by
\citet{PauldrachPuls90} when modeling the wind of P~Cygni. These findings
were generalized by \citet{Vink00, Vink01} in their work on stellar
wind models for OB-stars: In the intermediate/late O-star and early B-star
regime, the major contribution of driving lines in the lower wind (which are
responsible for initiating the mass-loss rate) is from \FeIV. Below \Teff =
23~kK, however, \FeIV\ recombines more or less abruptly to \FeIII. Since
\FeIII has more effective lines than \FeIV, $N_{\rm eff}$ increases (in
parallel with a decrease of $\alpha$, see also \citealt{Puls00}), which
leads to an increase in \Mdot\ and a decrease in the terminal velocity,
\vinf. This is the theoretical basis for the $\kappa$-effect discussed in
Sect.~2. \citet{Vink00} predict a typical increase in \Mdot\ by a factor of
five, and a decrease in \vinf\ by a factor of two. Consequently, the
wind-momenta from massive stars with \Teff $<$ 23~kK should be {\it higher}
than those from stars of earlier spectral types (see dashed/solid lines in
Fig.~\ref{fig6}.) This important prediction needs to be checked
observationally, not least because present day evolutionary codes
often incorporate the corresponding `mass-loss recipe'.

\noindent 
{\bf Test beds for the bi-stability jump (I): B[e]-supergiants?} 
The hybrid spectrum of B[e] supergiants can be explained by a two-component
wind, with an outflowing `disk' (equatorial wind) of low velocity, high
density and low ionization, and a high velocity, low density and highly
ionized polar wind \citep{Zickgraf86, Zickgraf89}. A first explanation of
this wind structure was given by \citet{LamersPauldrach91}, who combined the
effects of fast rotation and bi-stability jump, following the calculations
from \citet{PauldrachPuls90} for the latter: The high values of
$g_{\rm eff}$ together with the high ionization at the pole (\Teff\
calculated via von Zeipel) then give rise to a fast and thin polar wind,
whereas the low $g_{\rm eff}$/\Teff\ values at the equator induce a slow and
dense wind. (Note that \vinf\ scales with the photospheric escape velocity
and thus with $g_{\rm eff}$, see below).

Until now, the B[e]-sg mechanism is heavily debated. \citet{Owo98} pointed
out that \citet{LamersPauldrach91}, though accounting for gravity darkening
when calculating the ionization, did not include its impact on \Mdot\
(Eq.~\ref{mdot_rot}). When gravity darkening is included into the
accelerating flux, $F(\theta) \propto g_{\rm eff}(\theta)$, a `disk'
formation becomes almost impossible, due to the counteracting effects of
bi-stability and increased polar flux (see also \citealt{Puls08} and
references therein). Simulations by \citet{Pelupessy00}, on the other hand,
indicated that the bi-stability mechanism can work even when consistently
accounting for gravity darkening, at least for a density contrast up until
ten (observed values are on the order of hundred). \citet{Cure05} showed
that near critical rotation enables the wind to `switch' from the standard,
fast-accelerating solution to a slow, shallow-accelerating velocity
law. This, in combination with the bi-stability effect, can lead to the
formation of a slow and dense equatorial wind. \citet{Madura07}, finally,
confirmed and explained the `Cur\'e-effect', but argued that gravity
darkening is still a problem when aiming at a significant density contrast.

\noindent {\bf Test beds for the bi-stability jump (II): `normal'
B-supergiants.} Thus, it is still unclear whether B[e] supergiants can be
used to verify the bi-stability effect. Consequently, we now consider
`normal' B-supergiants. One of the predictions by \citet{Vink00} is an
abrupt decrease of \vinf (see also \citealt{Lamers95}) around the
bi-stability jump (around 23~kK). Let us first consider this effect.
Standard line-driven wind theory predicts that $\vinfe \approx 2.24
\alpha/(1-\alpha) \vesce$ (e.g., \citealt{Puls96}), and a compilation of
different measurements/analyses (mostly based on \citealt{Evans04} and
\citealt{Crowther06}) by \citet{Markova08} shows that the average ratio
\vinf/\vesc\ $\approx$ 3.3 for \Teff\ $>$ 23~kK and \vinf/\vesc\ $\approx$
1.3 for \Teff\ $<$ 18~kK, with a {\it gradual} decrease in between (see also
the original work by \citealt{Evans04, Crowther06}). Thus, there {\it is} an
effect on \vinf, but we also have to check the behaviour of the mass-loss
rates. Conventionally, this is done by plotting the modified wind-momentum
rate, $D_{\rm mom}$, as a function of the stellar luminosity, since one of
the major predictions from radiation driven wind theory is the well-known
wind-momentum luminosity relation (WLR, \citealt{WLR}), 
\[ 
\log D_{\rm mom} = \log  \bigl(\Mdote \vinfe (\Rstare/\Rsune)^{1/2}\bigr) 
\approx x \log(L/\Lsune) + \mbox{offset(spect. type, metallicity)}, 
\]
where $x$ has a similar dependence as the offset. (Theoretically, $x =
(\alpha - \delta)^{-1}$, where $\delta \approx 0.1$ accounts for ionization
effects.) 

Fig.~\ref{fig6} compares observationally inferred modified wind-momentum
rates for OB-super\-giants with the predictions from \citet{Vink00} (for
details, see \citealt{Markova08}). As pointed out above, the predicted WLR
for B-stars lies {\it above} the one for O-stars (more increase in \Mdot\
than decrease in \vinf), whereas the observations show the opposite. The
observed O-star rates (triangles, encircled in blue) lie above the
predictions, which can be explained by clumping effects (see below), whereas
the observed B-star rates for \Teff\ $<$ 22~kK lie well below the
predictions and those for \Teff\ $>$ 22~kK just connect
the O-star regime and the cooler B-stars. With respect to \Mdot\ itself, a
careful analysis shows that \Mdot\ either decreases in concert with \vinf
(more likely), or at least remains unaffected (less likely). Globally,
however, we do not see the predicted increase in \Mdot, though a certain
maximum around the location of the jump might be present \citep{Benaglia07}.
Thus, at least below the bi-stability jump there is a severe problem. Either
the predicted \Mdot\ for cooler objects are too high, or the `observed'
(i.e., derived) ones are too low. Accounting for the observed O-star rates,
the latter seems unlikely (and the inclusion of clumping would even
increase the discrepancy for the B-stars). A way out of the dilemma might be
the potential impact of the `slow' wind solution (see above) on
BA-supergiants, as suggested by Granada et al. (this volume).

\smallskip 
\noindent 
{\bf A separate population?} Returning to the problem of the low rotation
rates of B-supergiants and accounting for the above dilemma, one has to
admit that {\it if} indeed the mass-loss rates were not increasing at the
bi-stability jump, then there would be no bi-stability braking, and the
rapid drop of rotation below \Teff\ = 20~kK still needs to be explained. 
To this end, \citet{Vink10} discuss an alternative {\it scenario II} (see
also Langer, this volume): The cooler, slowly rotating supergiants might
form an entirely separate, non core hydrogen-burning population, e.g., they
might be products of binary evolution (though this is not generally 
expected to lead to slowly rotating stars), or they might be post-RSG or
blue-loop stars. 

Support of this second scenario is the finding that the majority of the
cooler objects (here: in the LMC) is {\it strongly Nitrogen-enriched}, which
was one of the outcomes of the {\sc vlt-flames} survey of massive stars
(Brott, this volume; see also \citealt{Evans08} for a brief summary of the
project). Vink et al. argue that ``although rotating models can in principle
account for large N abundances, the fact that such a large number of the
cooler objects is found to be N enriched suggests an evolved nature for
these stars.''

\begin{figure}[t]
\begin{center}
\begin{minipage}{12.8cm}
 \includegraphics[angle=180,width=12.8cm]{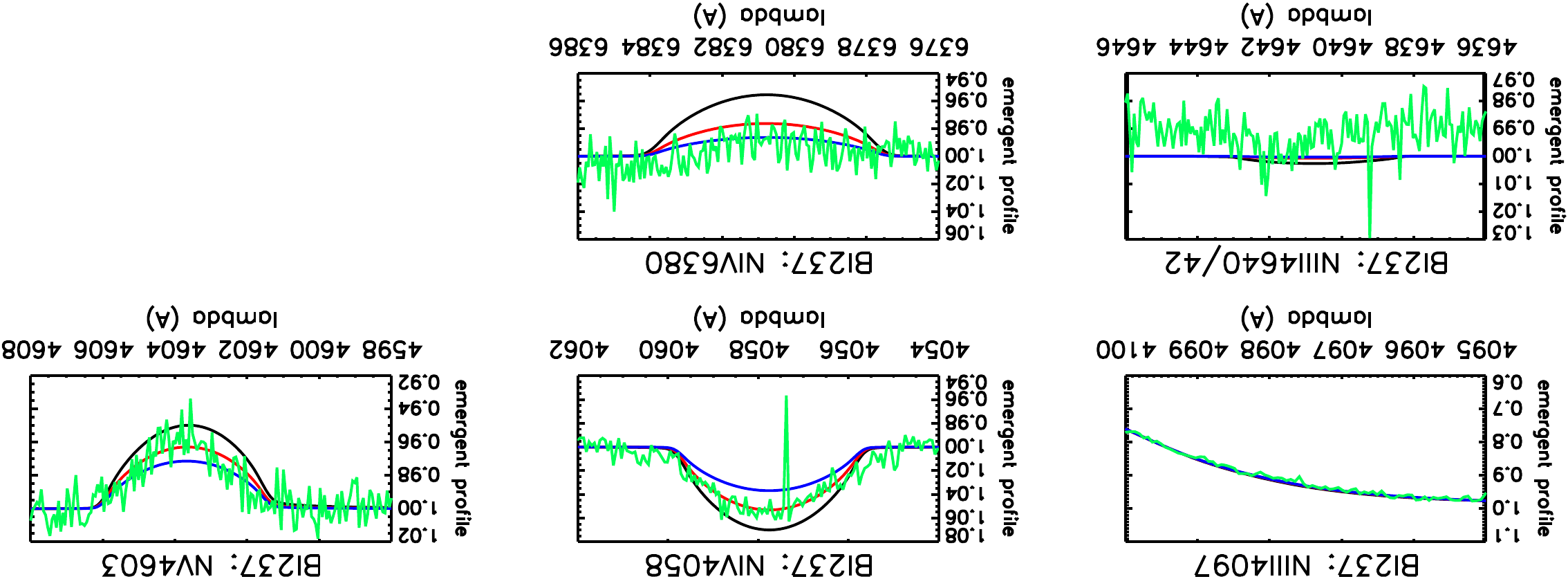} 
\end{minipage}
\begin{minipage}{12.8cm}
 \includegraphics[angle=180,width=12.8cm]{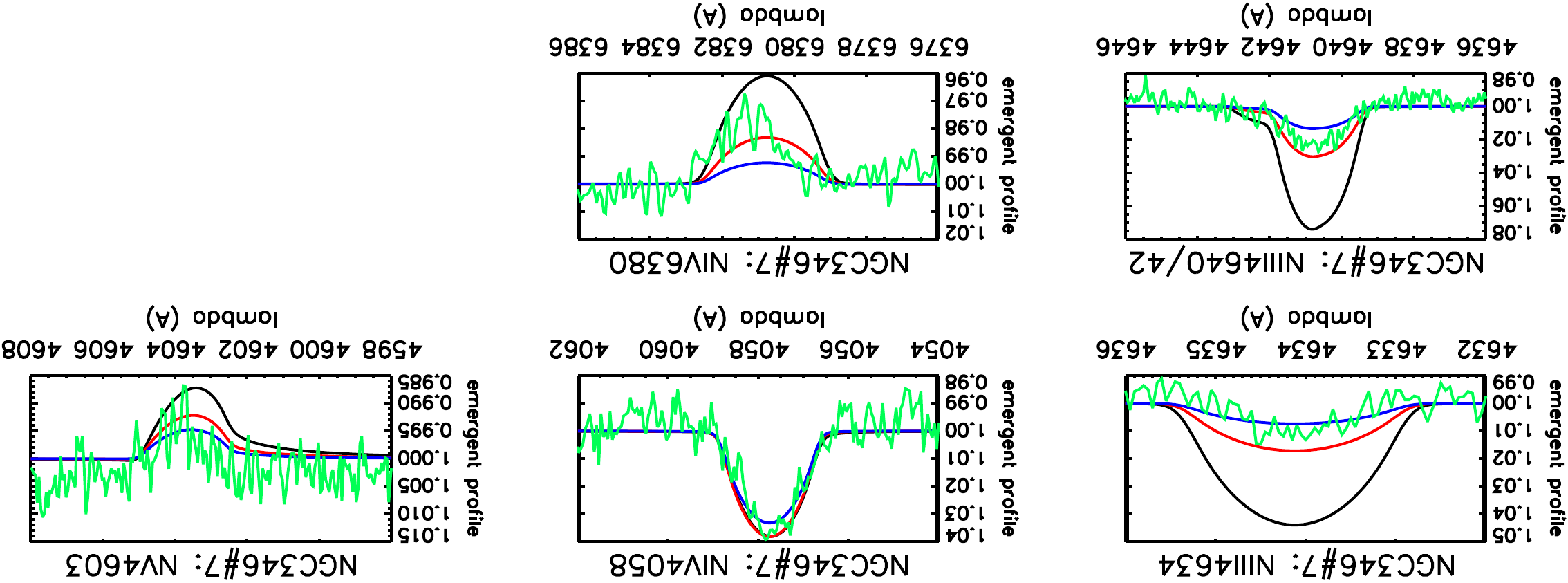} 
\end{minipage}
\caption{Strategic Nitrogen line profiles in the optical from ionization stages {\sc iii} to
{\sc v} for two early O-type stars. Observations in green, solid lines are
synthetic profiles (calculated by the atomospheric code {\sc fastwind}, 
\citealt{Puls05}) for different abundances: 
N/N$_\odot$=0.2 (blue), 0.4 (red) and 1.0 (black).
Upper panels: BI237 (O2V((f$^*$)) in the LMC), with \Teff\ =
52~kK and \logg\ = 4.0. The derived Nitrogen abundance is N/N$_\odot$ = 0.4 or
[N/H] = 7.38. Lower panels: NGC\#7 (O4V((f$^+$)) in the SMC), with \Teff\ =
45~kK and \logg\ = 4.0. The derived Nitrogen abundance is N/N$_\odot$ $\approx$
0.2{\ldots} 0.4 or [N/H]  $\approx$ 7.08{\ldots} 7.38.}
\label{fig7}
\end{center}
\end{figure}

\smallskip \noindent {\bf Nitrogen abundances from O-stars.} So far,
Nitrogen abundances could be derived only for a subset of the {\sc
vlt-flames} sample stars, and corresponding data are missing particularly
for the most massive and hottest stars. Indeed, when inspecting the
available literature for massive stars, one realizes that metallic
abundances, in particular of Nitrogen, which is {\it the} key element to
check evolutionary predictions, are scarcely found for O-type stars. The
simple reason is that they are difficult to determine, since the formation
of N{\sc iii}/{\sc iv} lines (and lines from similar ions of C and O) is
problematic due to the impact of various processes that are absent or
negligible at cooler spectral types, e.g., dielectronic recombination,
mass-loss, and clumping. Within the {\sc vlt-flames} project, progress is
under way \citep{Rivero10}, and in Fig.~\ref{fig7} we show two examples of
N-abundance determinations for two early type O-stars in the LMC and SMC.
Though no detailed comparison with evolutionary models has been made yet,
the derived abundances for both objects are consistent, within the
error-bars, with the average abundances from corresponding B-type stars of
early evolutionary stages, which are [N/H] = 7.13 $\pm 0.29$ for the LMC and
[N/H] = 7.24 $\pm 0.31$ for the SMC, respectively \citep{Hunter09}.  

\smallskip \noindent {\bf Wind clumping.} Mass loss is pivotal for the
evolution/fate of massive stars (e.g., the formation of GRBs critically
depends on the loss of angular momentum due to mass loss, see Ekstr\"om and 
Langer, this volume), their energy release, and their stellar yields. Thus,
reliable mass-loss rates are urgently required (ideally better than a factor
of two, \citealt{Meynet94}). O-star mass-loss rates derived from the
optical/radio have been found to be higher than predicted by the widely used
mass-loss recipe from \citet{Vink00} (see Fig.~\ref{fig6}). The present
hypothesis assumes that this discrepancy is due to neglected wind-clumping
(small scale density inhomogeneities), originating from the line-driven
instability, which results in overestimated mass-loss rates when using
recombination-based diagnostics \citep[ and references therein]{Puls08}. To
check and infer the effects due to optically thin and thick clumps, and due
to porosity in velocity space, on the various diagnostics,
\citet{Sundqvist10, Sundqvist11} have used the well observed star $\lambda$
Cep (O6I) as a {\bf test bed} to derive a mass-loss rate of 1.5\Mdu. This 
is a factor of four lower than corresponding `unclumped' values and a factor
of two lower than the predictions by \citet{Vink00}.

\section{Very brief summary and conclusions} We discussed OB-stars as
extreme condition {\bf test beds}, regarding effects due to pulsations, rapid
rotation, and mass-loss. Rapidly rotating B-supergiants (though scarce) are
particularly well suited to check a number of theoretical predictions, and 
the B1 supergiant $\gamma$~Ara may be a prime candidate for
future diagnostics.

\smallskip \noindent \footnotesize{{\bf Acknowledgements.} The authors
gratefully acknowledge a travel grant from the IAU and the local organizers
of this conference (J.P.), a grant from the IMPRS, Garching (J.O.S.),
and a research grant from the German DFG (J.G.R.G.).}

\begin{discussion}

\end{discussion}

\end{document}